 \documentclass[preprint2]{aastex}

\usepackage{graphicx}

\shorttitle{}
\shortauthors{Novakovi\' c et al.}

\begin{document}

\title{Asteroid secular dynamics: Ceres' fingerprint identified}

\author{Bojan~Novakovi\'c}
\affil{Department of Astronomy, Faculty of Mathematics, University of Belgrade, \\
Studentski trg 16, 11000 Belgrade, Serbia}
\email{bojan@matf.bg.ac.rs}

\author{Clara Maurel}
\affil{Institut Sup\' erieur de l'A\'eronautique et de l'Espace (ISAE-Supa\' ero),  \\
University of Toulouse,  \\ 10 avenue Edouard Belin, 31055 Toulouse Cedex
4, France}

\author{Georgios Tsirvoulis \& Zoran Kne\v zevi\' c}
\affil{Astronomical Observatory, Volgina 7, 11060 Belgrade 38, Serbia}

\begin{abstract}
Here we report on the significant role of a so far overlooked dynamical aspect, namely a
secular resonance between the dwarf planet Ceres and other asteroids. 
We demonstrate that this type of secular resonance can be the dominant dynamical factor in certain
regions of the main asteroid belt. 

Specifically, we performed a dynamical analysis of the asteroids belonging 
to the (1726) Hoffmeister family.
To identify which dynamical mechanisms are actually 
at work in this part of the main asteroid belt, i.e. to isolate
the main perturber(s), we study the evolution of this family in time.
The study is accomplished using numerical integrations of test particles
performed within different dynamical models. The obtained results reveal 
that the post-impact evolution of the Hoffmeister
asteroid family is a direct consequence of the nodal secular resonance with Ceres. 

This leads us to the conclusion that similar effects must exist in other parts of the
asteroid belt. In this respect, the obtained results shed light on an important and entirely 
new aspect of the long-term dynamics of small bodies. Ceres' fingerprint in asteroid dynamics, 
expressed through the discovered secular resonance effect, completely changes our 
understanding of the way in which perturbations by Ceres-like objects 
affect the orbits of nearby bodies.
\end{abstract}

\keywords{celestial mechanics --- minor planets, asteroids: general --- 
minor planets, asteroids: individual (Ceres)}

\section{Introduction}

Orbital resonances exist everywhere in the Solar System, and play an 
essential role in the dynamics of small bodies.
The synergy of fast and slow orbital angles produces a great assortment 
of resonant phenomena \citep{william1981,dermott1981,milani1994,nes1998}. 
Over the years numerous methods and models have been developed to interpret the complex 
dynamical environment of the main asteroid belt. It is well known that this 
region is sculpted by a web of mean-motion and secular resonances coupled with subtle 
non-gravitational forces \citep{gladman1997,farinella1999,bottke2006,minton2010,nov2010}. 
The implications of these effects on a large number of examples with unique dynamical characteristics 
have already been successfully described by existing dynamical models.
It is, however, still not possible to explain or predict all the dynamics of the main asteroid belt.

There are two general types of orbital resonances in the Solar system. 
The most intuitive
type, referred to as mean-motion resonances, occurs when the orbital periods of an 
asteroid and of a perturber are nearly commensurate. The second type, called secular resonance, 
concerns slowly varying angles like the longitude of perihelion or
the longitude of the ascending node. 

Secular resonances may play a significant role in the long-term dynamical stability
of a planetary system \citep{laskar1989,kne1991,michel1997}, however until now studies 
related to the dynamics of small solar system objects considered 
only planets as important perturbers.

The role of the most massive asteroids in the asteroid dynamics is generally assumed
to be small, and in most cases it is neglected. Still, it is known that perturbations
arising from the most massive asteroids could be important in some situations. Being
located relatively close to each other, the most important interaction between
the most massive and the rest of the asteroids intuitively occurs during their mutual close encounters.
The long-term effects of these encounters have been studied by many authors,
typically aiming to explain the evolution of asteroid families \citep{nes2002flora,carruba2003,nov2010lix,delisle2012,carruba2013}.
Moreover, it was demonstrated by \citet{christou2012} that there are populations of 
asteroids in the 1/1 mean motion resonances with the two most massive objects 
in the main belt, namely (1)~Ceres and (4)~Vesta.

Nevertheless, the importance of massive asteroids for secular dynamics and long-term
chaotic diffusion is generally accepted to be negligible; thus, it has never been studied.
In this paper, we show that this paradigm not only lacks justification, 
but it is actually incorrect. The results obtained here reveal that a nodal secular resonance with 
(1)~Ceres, namely $s-s_c$, plays a key role in the dynamics of asteroids belonging to 
the Hoffmeister family.

\section{Methods and results}

\begin{figure}
\centering
\includegraphics[angle=0,scale=.99]{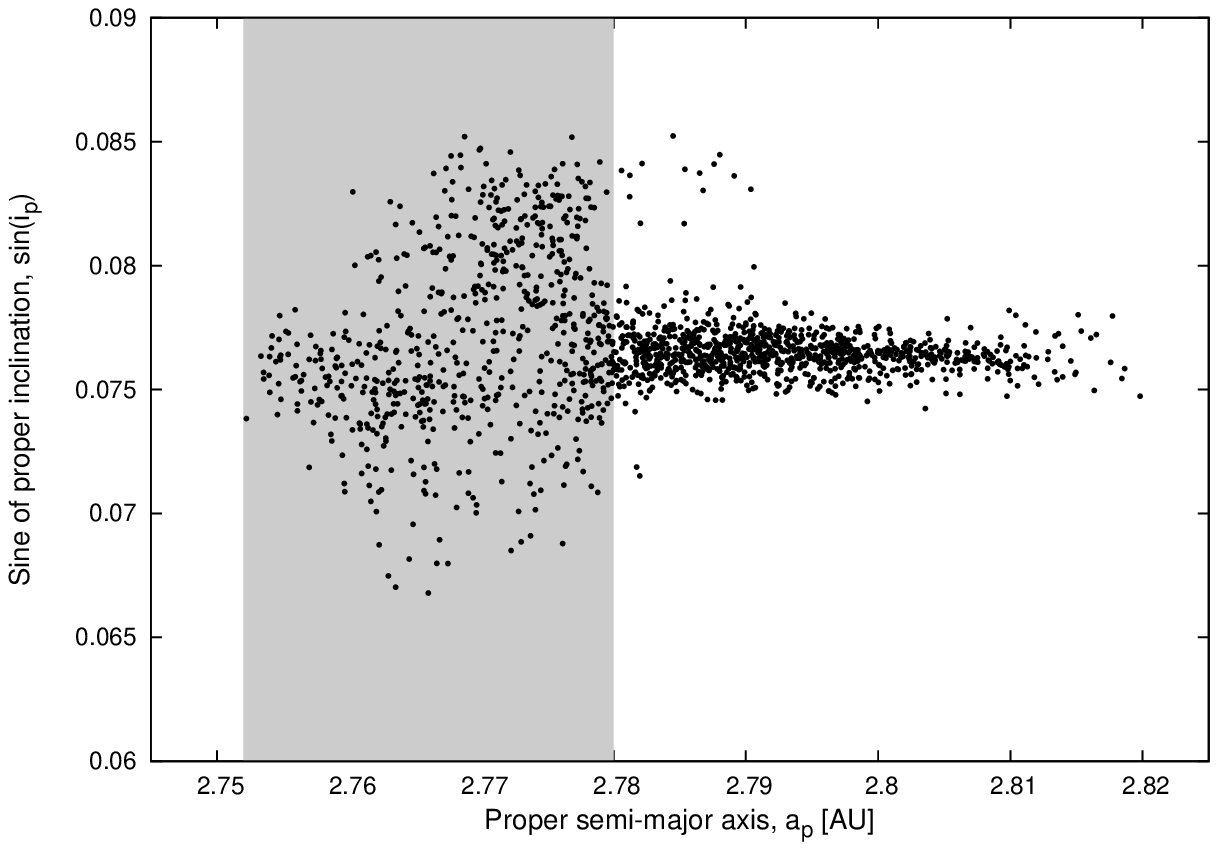}
\caption{•}
\label{fig1}
\end{figure}

The motivation for our work comes from the unusual shape of the Hoffmeister asteroid family \citep{milani2014} 
when projected on the proper orbital semi-major axis $a_{p}$ versus sine of proper orbital inclination $sin(i_{p})$ plane (Fig.~\ref{fig1}). 
The distribution of family members as seen in this plane clearly suggests different dynamical 
evolution for the two parts of the family delimited in terms of semi-major axis. 
The part located at $a_{p} < 2.78$~AU is dispersed, and seems to undergo significant evolution in 
inclination, contrary to that at $a_{p} > 2.78$~AU, which 
looks much more condensed and practically shows no similar evolution. 
Our goal here is to reveal the mechanism responsible for the observed asymmetry.

\hfill \break

\textbf{PLACE FIGURE 1 HERE.}

\hfill \break

Analyzing the region around the Hoffmeister family we found a few potentially important dynamical
mechanisms. The family is delimited in terms of semi-major axis by two mean motion resonances, the 3J-1S-1 
three body resonance with Jupiter and Saturn at 2.752~AU, and the 5/2J mean-motion resonance 
with Jupiter at 2.82~AU. 
Moreover, the region is crossed by the $z_1=g-g_6+s-s_6$ secular resonance, with $g$, $s$, $g_{6}$ and $s_{6}$ 
being the secular frequencies of the asteroid's and Saturn's orbits.

\subsection{Numerical simulations}

\subsubsection{Dynamical model}

To identify which ones, if any, among the different possible dynamical mechanisms are actually at work
here, we performed a set of numerical integrations. For this purpose we employed the
\emph{ORBIT9} integrator embedded in the multipurpose \emph{OrbFit} package\footnote{Available from http://adams.dm.unipi.it/orbfit/}.
The dynamical model includes the gravitational effects of the Sun and the four outer planets, from Jupiter
to Neptune. It also accounts for the Yarkovsky thermal effect, a subtle non-gravitational
force due to the recoil force of anisotropically emitted thermal radiation by a rotating body \citep{bottke2006}, causing mainly a secular drift in semi-major axis. The indirect effect of the 
inner planets is accounted for by applying a barycentric
correction to the initial conditions.

Our simulations follow the long-term orbital evolution of test particles initially distributed randomly
inside an ellipse determined by the Gauss equations. This ellipse corresponds
to the dispersion of the Hoffmeister family members immediately after the breakup event,
assuming an isotropic ejection of the fragments from the parent body.
 
The total number of particles used is 1678, the same as the number of asteroids 
we currently identified as members of the family. The family membership is determined
utilizing the hierarchical clustering method and standard metric as proposed by
\citet{hcm1990}.

For simplicity, the Yarkovsky effect is approximated in terms of a pure
along-track acceleration, inducing on average the same semi-major axis drift
speed $da/dt$ as predicted from theory.\footnote{This model of the net Yarkovsky 
force is a reasonable approximation over short timescales, 
but may not be accurate enough in the long term, because the spin axis or the rotational 
period may change. The spin evolution of an asteroid not subject to collisions, is
expected to be dominated by the Yarkovsky-O'Keefe-Radzievskii-Paddack (YORP) 
effect \citep[see e.g.][]{Rubincam2000}.
The models predict that YORP torques may evolve bodies toward asymptotic rotational states 
\citep{CapVok2004}, or could cause reshaping that would significantly increase 
the time over which objects can preserve their sense of rotation \citep{Cotto-Figueroa2015}. 
This has implications for the Yarkovsky 
effect, however, the constant Yarkovsky drift we used here represents the long-term 
average of this effect. Regardless of the actual behavior of any single body, the average 
drift rate should be nearly constant for a large enough statistical sample.} Assuming an isotropic distribution 
of spin axes in space, to each particle we randomly assign
a value from the interval $\pm (da/dt)_{max}$, where $(da/dt)_{max}$ is the estimated 
maximum of the semi-major axis drift speed due to the Yarkovsky force. 
The value of $(da/dt)_{max}$ is determined using
a model of the Yarkovsky effect developed by \citet{vok1998,vok1999}, and
assuming thermal parameters appropriate for regolith-covered $C$-type objects.
In particular, we adopt values of $\rho_{s}$ = $\rho_{b}$ = 1300~$kg~m^{-3}$
for the surface and bulk densities \citep{carry2012}, $\Gamma$ = 250~$J~m^{-2}~s^{-1/2}~K^{-1}$
for the surface thermal inertia \citep{delbo2009}, and $\epsilon$ = 0.95 
for the thermal emissivity parameter.
In this way we found that for a body of $D=1$~km in diameter $(da/dt)_{max}$
is about $4 \times 10^{-4}$~AU/Myr.
Next, we select sizes of the test particles equal to sizes of the Hoffmeister 
family members, estimated using their absolute magnitudes provided by AstDys database\footnote{http://hamilton.dm.unipi.it/astdys/}, 
and geometric albedo of $p_{v}$ = 0.047 \citep{wise}. Finally, since Yarkovsky effect scales as $ \propto 1/D$, the particle sizes are then used to calculate corresponding 
value of $(da/dt)_{max}$ for each particle, by scaling from the reference value 
for objects of $D=1$~km.

The orbits of the test particles are propagated for 300~Myr, which is a rough estimate of the age 
of the Hoffmeister family \citep{nes2005,spoto2015}. Time series of mean orbital 
elements\footnote{The mean orbital elements are obtained by removal of the short-periodic perturbations
from the instantaneous osculating elements.} are produced using on-line
digital filtering \citep{carpino1987}. Then, for each particle we compute the synthetic 
proper elements \citep{knemil2000} for consecutive intervals of 10~Myr. This allows us to study 
the evolution of the family in the space of proper orbital elements.

If our first dynamical model were complete we should be able to
reproduce the current shape of the family. 
However, we found that the shape cannot be 
reproduced with the afore described model. 
In particular, in these simulations we observed 
only a dispersion of the semi-major axis caused by the Yarkovsky effect, but no evolution
in inclination (Fig.~\ref{fig2}).
This implies that neither mean-motion nor secular resonances involving the major outer planets 
are responsible for the strange shape of the Hoffmeister family. 

\hfill \break

\textbf{PLACE FIGURE 2 HERE.}

\hfill \break

In order to clarify the situation we turned our attention to the inner
planets, and their possible role in the evolution of the family. 
For a representative sample of about 200 test particles we repeated the above 
described simulations using a model with seven planets (from Venus
to Neptune), which also includes the Yarkovsky thermal force.
However, the 7-planet model did not give any different results, which remain practically 
the same as the one obtained within the model with the four outer planets only.
Thus, obviously there is still something missing in the dynamical model.

\subsubsection{Extended dynamical model}

Being left with almost no other option, 
we turn our attention to asteroid (1)~Ceres. Having a proper orbital semi-major axis 
of 2.767~AU, and therefore being inside the range covered by the family members, 
it seems to be the only remaining candidate. 
Hence, we again numerically 
integrated the same test particles, using the 4-planet dynamical model but this time also including
Ceres as a perturbing body\footnote{In these simulations, for mass of Ceres we used a value of 
$4.757 \times 10^{-10} M_{\odot}$, as estimated by \citet{baer}.}. 

These new simulations already after about 150~Myr very nearly matched the current
spreading of the family in the semi-major axis versus inclination plane, 
clearly implicating Ceres as the culprit for what we see today.
The striking feature observed in these runs is a fast dispersion of orbital inclinations for 
objects with semi-major axis of about 2.78~AU. After 15~Myr of evolution the spread in sine of inclination
is 3 times larger than the initial one, as can be seen in the middle panel in Fig.~\ref{fig2}.

\subsubsection{Mechanism of Ceres perturbations}

The fact that the main perturber is Ceres raises a very important question. 
What is the exact mechanism by which Ceres is perturbing
the members of the Hoffmeister family to such a high degree? The current paradigm suggests
this may be the result of close encounters, or it might be the consequence of the 1/1 mean motion resonance 
with Ceres. However, our simulations undoubtedly show that none of these two mechanisms could explain
the evolution of the family. Actually, we found that most of the evolution is taking place within a
narrow range of the semi-major axis. However, this range does not correspond to the location of the
1/1 resonance with Ceres, neither is there any reason for close encounters to affect only
objects within this specific range of semi-major axes. Moreover, it is very unlikely that these two
effects would primarily affect orbital inclinations. Finally, regardless of the mechanism,
such a large perturbation on asteroids caused by Ceres has never been observed in the asteroid belt. 
This situation motivates us to test 
some other mechanisms which are at work in this region, despite generally being accepted to be negligible.
These are the secular resonances with asteroid Ceres. 

Analyzing the secular frequencies
of the Hoffmeister family members we immediately notice that some of these are very
close to the nodal frequency of Ceres ($s_c=-59.17$~arcsec/yr). This is an interesting 
fact because the secular resonances involving the nodal frequencies are known to affect 
mainly the orbital inclination. 

To better understand the reasons for the dispersion in inclination, and to determine the possible
role of the secular resonance with Ceres, we pick a few particles
that experienced significant changes in inclination during our numerical simulations, and
analyzed their behavior in more detail. This analysis identified a mechanism responsible 
for the evolution of orbital inclinations, revealing a completely new role
of Ceres in asteroid dynamics. 

As an illustration, in Fig.~\ref{fig3} we show the evolution of one
test particle. This particle was initially located at a semi-major axis of 2.788~AU, with
its Yarkovsky induced drift set to be negative, forcing it to move towards
the Sun. After about 65~Myr this particle enters the region where the fast dispersion in orbital
inclination has been observed. During the time spent in this area, the particle's
inclination has experienced a fast increase, with the average value jumping from about 
4.4 to 4.9 degrees. 

\hfill \break

\textbf{PLACE FIGURE 3 HERE.}

\hfill \break

\begin{figure}
\centering
\includegraphics[angle=0,scale=.82]{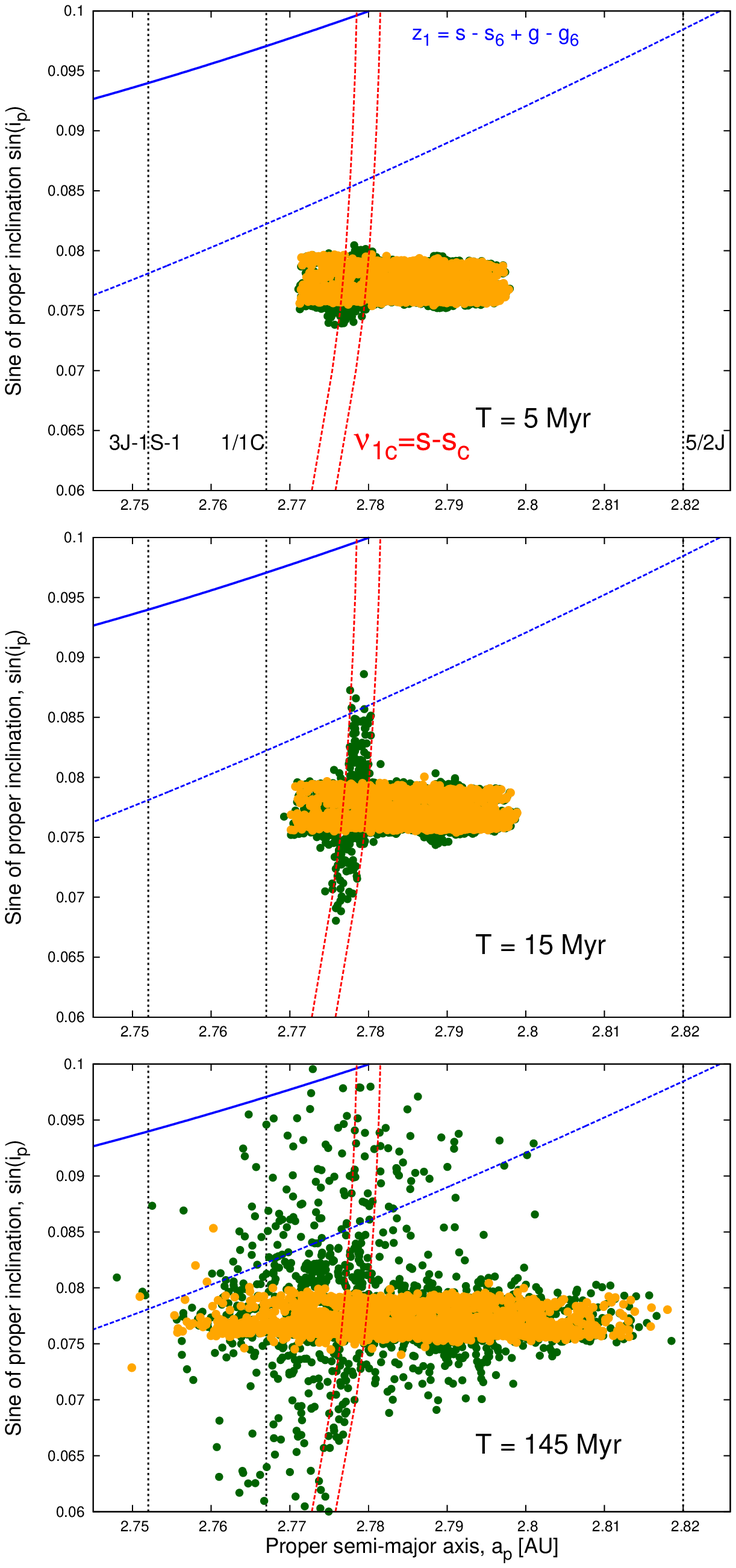}
\caption{•}
\label{fig2}
\end{figure}

Let us recall here that a resonance occurs when the corresponding critical angle 
librates\footnote{A libration is the oscillation of an angle around a fixed point, 
contrary to the circulation when the angle cycles over all values from 0 to 360 degrees.}. 
In the case of the $s-s_c$ secular resonance the critical angle is
$\sigma=\Omega-\Omega_c$, where $\Omega$ and $\Omega_c$ are the longitude of the ascending node 
of an asteroid and Ceres, respectively. 
Clearly, as can be seen in Fig.~\ref{fig3}, the period of increase in inclination exactly
corresponds to the period of libration of the critical angle of the $\nu_{1c} = s-s_c$ secular resonance.
This correlation is a direct proof that this resonance is responsible for the
evolution and observed spread in the orbital inclination for asteroids belonging to
the Hoffmeister asteroid family. In Fig.~\ref{fig2} we plotted the location of 
the $\nu_{1c}$ resonance.

Though our numerical simulations clearly show that passing through the $\nu_{1c}$ resonance may
cause significant changes in orbital inclination, we further investigated the mechanism. 
A key to understand the observed behavior are the cyclic oscillations in inclination for objects trapped 
inside this resonance (see Fig.~\ref{fig4}). 
In the scenario with the Yarkovsky effect included in the model, some of the objects
are reaching the border of the $\nu_{1c}$ while steadily drifting in semi-major axis,
and enter it at random value of the inclination cycle. During the time spent inside the
resonance their inclination is continuously repeating the cycles. However, as the semi-major axis is
evolving due to the Yarkovsky effect, the objects must sooner or later reach 
the other border of the $\nu_{1c}$, and subsequently exit from the resonance. As the exit also 
happens at a random value of the inclination cycle, the values of orbital inclination
with which bodies enter the cycle typically differ from those with which they exit.
In this way the $\nu_{1c}$ resonance changes the inclination of objects that cross it. 
Certainly, this mechanism would not work without the Yarkovsky effect.

Interestingly, this process is very similar
to the one observed inside another relatively week secular resonance, namely the $g+2g_{5}-3g_{6}$, 
which affects members of the Koronis asteroid family \citep{bottke2001}.

\hfill \break

\textbf{PLACE FIGURE 4 HERE.}

\hfill \break

It is also worth mentioning that although passage across the $\nu_{1c}$ secular 
resonance could result in a very fast change in orbital inclination, the total changes 
are limited and could not exceed the maximal variations. The amplitude of variations in 
orbital inclination is not the same for all objects, but it is generally similar as the one 
shown in Fig.~\ref{fig4}.

Moreover, we have found that our test particles
typically spend 15-30~Myr inside the $\nu_{1c}$, before being moved outside by the Yarkovsky effect.
Thus, as a libration period is very long (about 40~Myr), most of the particles spend less than
one librating cycle inside this resonance (Fig.~\ref{fig3}).

Finally, let us re-examine the role of $z_{1}$ for the members of
the Hoffmeister family. The results presented above clearly indicate that without
Ceres in the model, the $z_{1}$ does not affect the family. Still, with Ceres
included in the dynamical model $z_{1}$ may only affect a limited number of members.
Our analysis has shown that about 5 per cent of family members could reach this resonance
after their inclination is pumped-up, at least a bit, by the $\nu_{1c}$ resonance (see Fig.~\ref{fig2}). 
The orbital inclination of these objects is then additionally dispersed by $z_{1}$,
being the main reason for the slightly different distributions towards high and low inclinations.
Moreover, although the analysis along this line is beyond the scope of the letter,
we noticed that the $z_{1}$ resonance also slightly affects the orbital eccentricity
of family members.

Nevertheless, the role of the $z_{1}$ in the dynamical evolution of
the Hoffmaister family members is minor compared to the role of the $\nu_{1c}$ 
secular resonance. Thus, the evolution of the family is almost completely determined
by the combined effect of the $\nu_{1c}$ resonance and the Yarkovsky effect, and would be 
practically the same even if the $z_{1}$ is not that close.

\section{Conclusions}

This is the first time a compelling evidence for orbital evolution of small bodies
caused by a secular resonance with an asteroid has been found. 
We prove that the post-impact transformation of the Hoffmeister
asteroid family is a direct consequence of the nodal secular resonance with Ceres. This result 
has very important repercussions for our view of how Ceres-size bodies affect the dynamics of 
nearby objects, and opens new possibilities to study such effects in the main asteroid belt 
and beyond. Examples include the dynamics of specific asteroid populations, the early phases 
of planetary formation, and extra-solar debris disks.

\hfill \break

\acknowledgments

The authors would like to thank Aaron Rosengren for his comments and 
suggestions on the manuscript. This work has been supported by the 
European Union [FP7/2007-2013], project: STARDUST-The Asteroid and 
Space Debris Network. BN and ZK also acknowledge support by the Ministry 
of Education, Science and Technological Development of the Republic 
of Serbia, Project 176011. Numerical simulations were run on the PARADOX-III cluster
hosted by the Scientific Computing Laboratory of the Institute of Physics Belgrade.

\clearpage

\begin{center}
\textbf{FIGURE CAPTIONS:}
\end{center}

\textbf{Fig. 1:} The Hoffmeister family in the space of proper orbital semi-major axis \textit{versus} 
sine of proper orbital inclination. Note the strange shape of the family in this plane, in particular the 
large dispersion in the sin($i_p$) direction of the part located at semi-major axis less than about 2.78~AU (denoted by a shaded area in this figure).

\textbf{Fig. 2:} The evolution of the Hoffmeister family in the space of proper orbital elements.
The three panels show the distribution of the test particles after 5, 15, and 145~Myr of the evolution, 
from top to bottom, respectively. The orange dots represent the evolution of the particles 
within the dynamical model that includes the four giant planets, from Jupiter to Neptune, and accounts 
for the Yarkovsky effect. The dark-green dots show the evolution when Ceres is added to the previous model. 
The vertical dashed lines mark the locations of mean motion resonances.
The inclined blue lines denote the position of the $z_{1}$ secular resonance; the solid line refers 
to the center, and the dashed line refers to the approximate lower border of this resonance. 
Finally, the red dashed lines mark the approximate 
position of the $\nu_{1c}=s-s_{c}$ secular resonance with Ceres. These plots clearly show
very different dynamical evolutions of particles, depending whether or not Ceres is considered as a perturbing body. 
While after 5~Myr the two distributions are still quite similar, after 15~Myr a remarkable difference 
is visible, with green dots evolving along the $\nu_{1c}$ resonance. 
The last snapshot corresponds to the distribution of the green dots
after 145~Myr, very similar to the distribution of the real Hoffmeister's family members, indicating 
that this resonance is responsible for the strange shape of the family.

\textbf{Fig. 3:} Time evolution of the critical angle of the $\nu_{1c}=s-s_{c}$ secular resonance (top), 
and the mean inclination (bottom) for one of the test particles. The correspondence of the 
time periods in which the critical angle is librating and the mean inclination is rapidly increasing 
(marked in both panels by the shaded area) clearly reveals the importance of the $\nu_{1c}$ secular 
resonance with Ceres in asteroid dynamics in this part of the main asteroid belt. 
The solid black line shows the average of the mean inclination to better appreciate the evolution.

\begin{figure}
\centering
\includegraphics[angle=0,scale=.99]{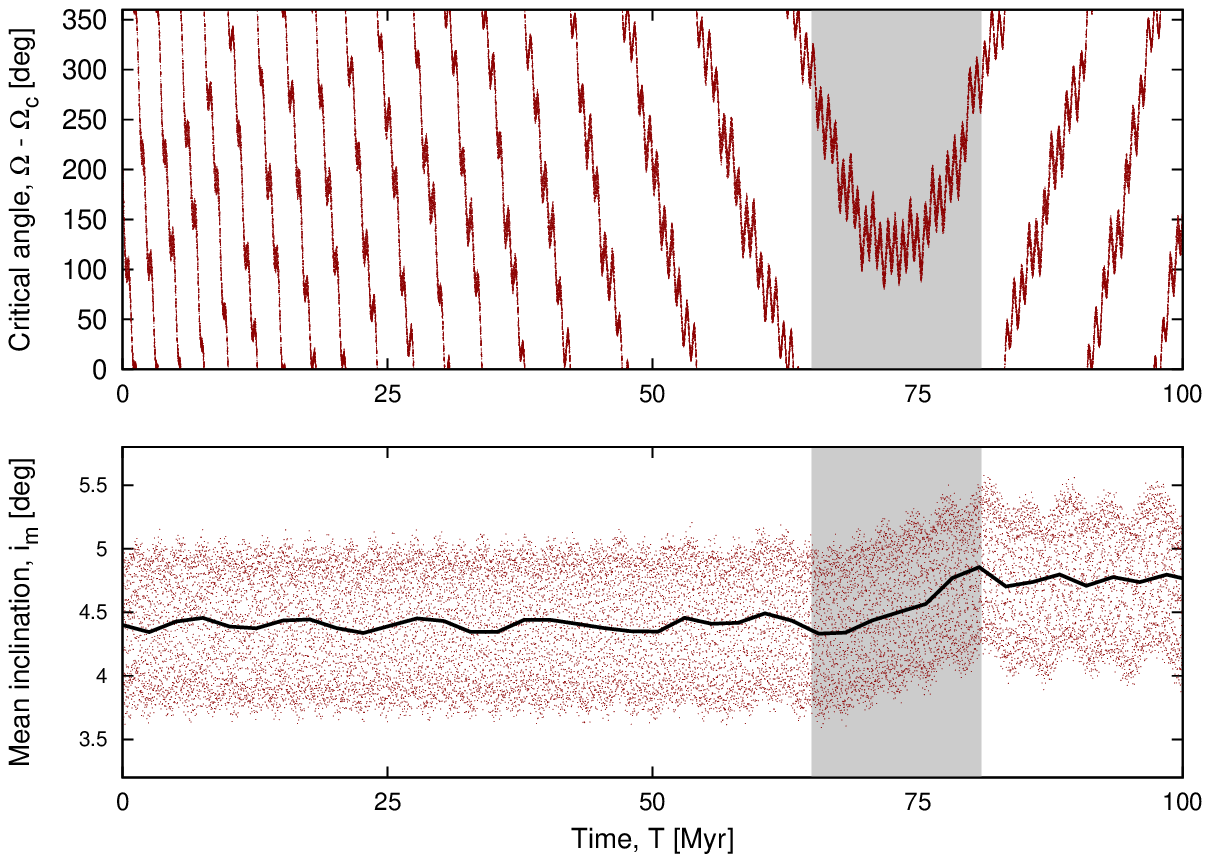}
\caption{•}
\label{fig3}
\end{figure}

\begin{figure}
\centering
\includegraphics[angle=-90,scale=.49]{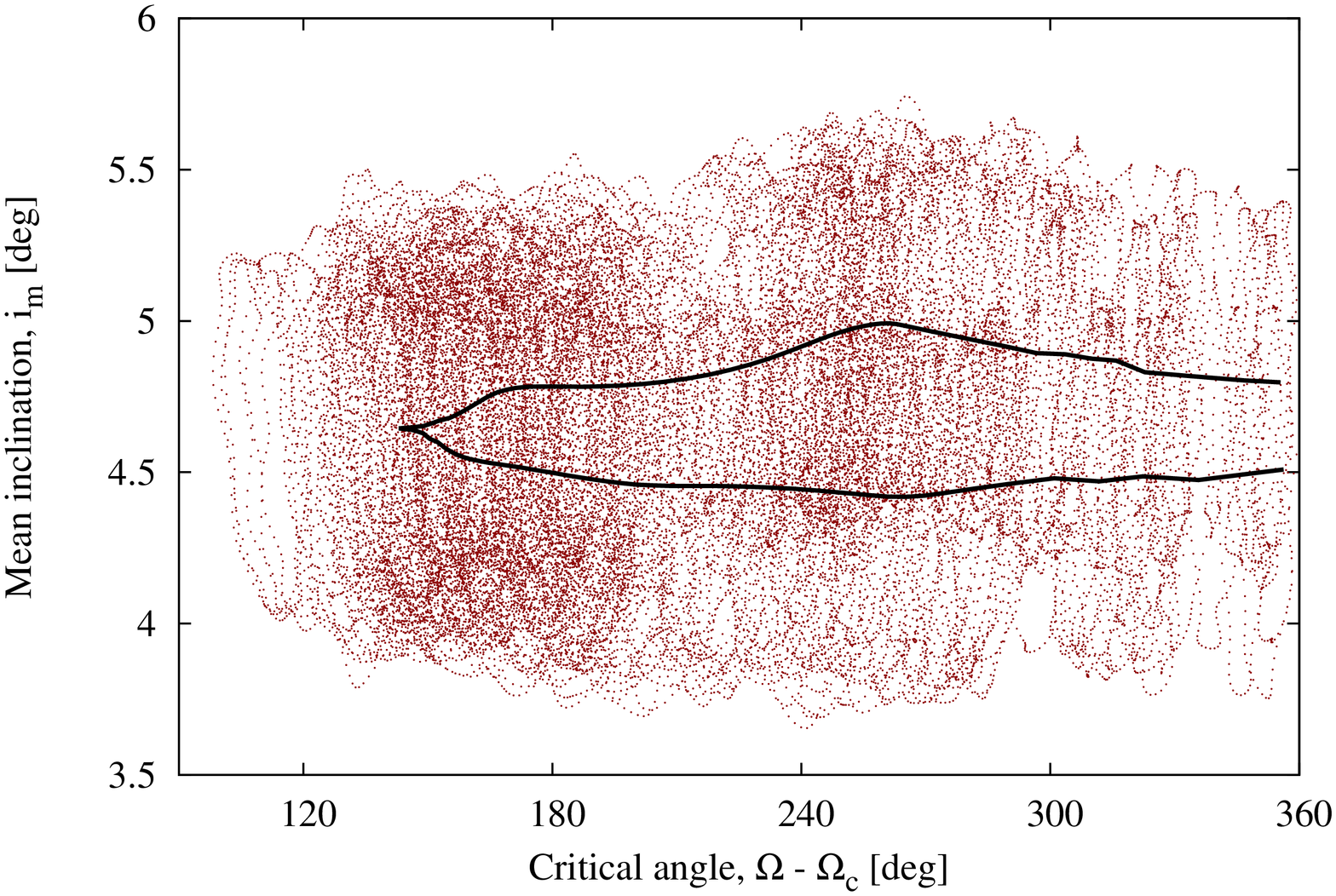}
\caption{•}
\label{fig4}
\end{figure}

\textbf{Fig. 4:} Variation of the mean orbital inclination $i_m$ vs. the critical angle
$\Omega - \Omega_c$ for the same test particle as shown in Fig.~\ref{fig3}..
Note, however, that only time interval when this particle is inside the resonance 
is shown, i.e. from about 66 to 85 Myr. During this time span
the inclination of the particle undergoes cyclic variations with two
different periods. The first mode of these oscillations (red points) has
a very short period of about 38 kyr; if averaged out, it reveals the
second, long period one (denoted by the black line), with a period of
about 25 Myr. Note that the long period is actually associated to the
libration of the critical angle. A similar situation was also observed
for secular resonances with the major planets \citep[see e.g.][]{froeschle1991}.

\newpage

\clearpage

\newpage

\clearpage

\newpage

\newpage

\begin{thebibliography}{}

\bibitem[Baer et al.(2011)]{baer} Baer, J., Chesley, S.~R., 
\& Matson, R.~D.\ 2011, \aj, 141, 143 

\bibitem[Bottke et al.(2001)]{bottke2001} Bottke, W.~F., 
Vokrouhlick{\'y}, D., Broz, M., Nesvorn{\'y}, D., 
\& Morbidelli, A.\ 2001, Science, 294, 1693 

\bibitem[Bottke et al.(2006)]{bottke2006} Bottke, W.~F., 
Vokrouhlick{\'y}, D., Rubincam, D.~P., \& Nesvorn{\'y}, D.\ 2006, 
Annual Review of Earth and Planetary Sciences, 34, 157

\bibitem[{\v C}apek \& Vokrouhlick{\'y}(2004)]{CapVok2004} 
{\v C}apek, D., \& Vokrouhlick{\'y}, D.\ 2004, \icarus, 172, 526

\bibitem[Carpino et al.(1987)]{carpino1987} Carpino, M., Milani, A., \& Nobili, A.~M.\ 1987, \aap, 181, 182

\bibitem[Carruba et al.(2003)]{carruba2003} Carruba, V., Burns, 
J.~A., Bottke, W., \& Nesvorn{\'y}, D.\ 2003, \icarus, 162, 308

\bibitem[Carruba et al.(2013)]{carruba2013} 
Carruba, V., Huaman, M., Domingos, R.~C., \& Roig, F.\ 2013, \aap, 550, A85 

\bibitem[Carry(2012)]{carry2012} Carry, B.\ 2012, \planss, 73, 98 

\bibitem[Christou \& Wiegert(2012)]{christou2012} Christou, A.~A., \& Wiegert, P.\ 2012, \icarus, 217, 27 

\bibitem[Cotto-Figueroa et al.(2015)]{Cotto-Figueroa2015} Cotto-Figueroa, 
D., Statler, T.~S., Richardson, D.~C., \& Tanga, P.\ 2015, \apj, 803, 25

\bibitem[Delb\' o \& Tanga(2009)]{delbo2009} Delb\' o, M., \& Tanga, P.\ 2009, \planss, 57, 259 

\bibitem[Delisle \& Laskar(2012)]{delisle2012} 
Delisle, J.-B., \& Laskar, J.\ 2012, \aap, 540, AA118 

\bibitem[Dermott \& Murray(1981)]{dermott1981} Dermott, S.~F., \& Murray, C.~D.\ 1981, \nat, 290, 664 

\bibitem[Farinella \& Vokrouhlick{\'y}(1999)]{farinella1999} 
Farinella, P., \& Vokrouhlick{\'y}, D.\ 1999, Science, 283, 1507

\bibitem[Froeschl{\'e} et al.(1991)]{froeschle1991} Froeschl{\'e}, C., Morbidelli, A., \& Scholl, H.\ 1991, \aap, 249, 553 

\bibitem[Gladman et al.(1997)]{gladman1997} Gladman, B.~J., 
Migliorini, F., Morbidelli, A., et al.\ 1997, Science, 277, 197

\bibitem[Kne{\v z}evi{\'c} et al.(1991)]{kne1991} Kne{\v z}evi{\'c}, Z., Milani, 
A., Farinella, P., Froeschl{\'e}, C., \& Froeschl{\'e}, C.\ 1991, \icarus, 93, 316

\bibitem[Kne{\v z}evi{\'c} \& Milani(2000)]{knemil2000} 
Kne{\v z}evi{\'c}, Z., \& Milani, A.\ 2000, Celestial Mechanics and Dynamical Astronomy, 78, 17

\bibitem[Laskar(1989)]{laskar1989} Laskar, J.\ 1989, \nat, 338, 
237

\bibitem[Masiero et al.(2011)]{wise} Masiero, J.~R., 
Mainzer, A.~K., Grav, T., et al.\ 2011, \apj, 741, 68 

\bibitem[Milani \& Knezevic(1994)]{milani1994} Milani, A., \& Kne{\v z}evi{\'c}, Z.\ 1994, \icarus, 107, 219

\bibitem[Milani et al.(2014)]{milani2014} Milani, A., Cellino, A., 
Kne{\v z}evi{\'c}, Z., et al.\ 2014, \icarus, 239, 46 

\bibitem[Michel \& Froeschl{\'e}(1997)]{michel1997} Michel, P., \& Froeschl{\'e}, C.\ 1997, \icarus, 128, 230 

\bibitem[Minton \& Malhotra(2010)]{minton2010} Minton, D.~A., \& Malhotra, R.\ 2010, \icarus, 207, 744 

\bibitem[Nesvorn{\'y} \& Morbidelli(1998)]{nes1998} 
Nesvorn{\'y}, D., \& Morbidelli, A.\ 1998, \aj, 116, 3029 

\bibitem[Nesvorn{\'y} et al.(2005)]{nes2005} Nesvorn{\'y}, D., 
Jedicke, R., Whiteley, R.~J., \& Ivezi{\'c}, {\v Z}.\ 2005, \icarus, 173, 132 

\bibitem[Nesvorn{\'y} et al.(2002)]{nes2002flora} Nesvorn{\'y}, D., 
Morbidelli, A., Vokrouhlick{\'y}, D., Bottke, W.~F., 
\& Bro{\v z}, M.\ 2002, \icarus, 157, 155 

\bibitem[Novakovi{\'c} et al.(2010)]{nov2010lix} Novakovi{\'c}, 
B., Tsiganis, K., \& Kne{\v z}evi{\'c}, Z.\ 2010, Celestial Mechanics and Dynamical Astronomy, 107, 35 

\bibitem[Novakovi{\'c} et al.(2010)]{nov2010} Novakovi{\'c}, 
B., Tsiganis, K., \& Kne{\v z}evi{\'c}, Z.\ 2010, \mnras, 402, 1263 

\bibitem[Rubincam(2000)]{Rubincam2000} Rubincam, D.~P.\ 2000, \icarus, 148, 2 

\bibitem[Spoto et al.(2015)]{spoto2015} Spoto, F., Milani, A., 
\& Knezevic, Z.\ 2015, arXiv:1504.05461

\bibitem[Vokrouhlick{\'y}(1998)]{vok1998} Vokrouhlick{\'y}, D.\ 1998, \aap, 335, 1093

\bibitem[Vokrouhlick{\'y}(1999)]{vok1999} Vokrouhlick{\'y}, D.\ 1999, \aap, 344, 362 

\bibitem[Williams \& Faulkner(1981)]{william1981} 
Williams, J.~G., \& Faulkner, J.  \ 1981, \icarus, 46, 390

\bibitem[Zappala et al.(1990)]{hcm1990} Zappala, V., Cellino, 
A., Farinella, P., \& Kne{\v z}evi{\'c}, Z.\ 1990, \aj, 100, 2030 

\end{thebibliography}
\end{document}